\documentclass{acmconf}
\usepackage{code,url}

\mathchardef\col="003A  
\mathchardef\semi="603B 
\mathchardef\bang="6021 

\mathchardef\lt="313C  
\mathchardef\gt="313E  

\newdimen\cascadeindent
\newdimen\cascdimen
\newcommand{\cascitem}{\\\global\advance\cascdimen by\cascadeindent\hspace{\cascdimen}}

\newcommand{\cascback}[1]{\\\global\advance\cascdimen by-#1.0\cascadeindent\hspace{\cascdimen}}

\newcommand{\bigleft}{\left(\begin{array}{@{}l@{}}}
\newcommand{\bigright}{\end{array}\right)}

\chardef\tilde="7E
\chardef\caret="5E
\chardef\lbrace="7B
\chardef\rbrace="7D

\newcommand{\figfoot}{\vspace{1ex}\hrule}
\newcommand{\fighead}{\hrule\vspace{1.5ex}}

\newcommand{\z}{\mbox{}}

\title{Hygienic Source-Code Generation Using Functors}
\author{Karl Crary}
\affiliation{Carnegie Mellon University}

\begin{document}

\maketitle{}

\begin{abstract}
Existing source-code-generating tools such as Lex and Yacc suffer from
practical inconveniences because they use disembodied code to
implement actions.  To prevent this problem, such tools could generate
closed functors that are then instantiated by the programmer with
appropriate action code.  This results in all code being type checked
in its appropriate context, and it assists the type checker in
localizing errors correctly.  We have implemented a lexer generator
and parser generator based on this technique for Standard ML, OCaml, and
Haskell.
\end{abstract}

\section{Introduction}
\label{sec:intro}

Compiler implementers have a love-hate relationship with
source-code-generating tools such as Lex~\cite{lesk:lex} (which generates lexers from
regular expressions) and Yacc~\cite{johnson:yacc} (which generates shift-reduce
parsers from context-free grammars).  These tools automate
the some of the most tedious parts of implementing a parser, but they
can be awkward to use.

One of the main awkward aspects of such tools is the {\em disembodied
code problem.}  To build a lexer or a parser, these tools cobble
together snippets of code (each implementing an action of the
lexer/parser) supplied by the programmer in a lexer/parser
specification file.  Unfortunately, the code snippets, as they appear
in the specification file, are divorced from their ultimate context.
The tools manipulate them as simple strings.\footnote{Such strings may
even include syntax errors, which are duly copied into the output code.
Typically the tool does not even ensure that delimiters are matched.}

This makes programming awkward in several ways.  Functions and other
values are passed into the snippets using identifiers that are bound
nowhere in the programmer's code, nor even introduced by a
pseudo-binding such as \textcd{open}.  Rather, the snippet is copied
into a context in which such identifiers are in scope.  This can make
code difficult to read.

More importantly, disembodied code makes debugging challenging,
because the code seen by the compiler bears little resemblance to the
code written by the programmer.  For example, consider the following
line from an ML-Lex~\cite{appel+:ml-lex} specification:

\codebegin
{whitespace}+ => ( lex () );
\codeend

\noindent
This line tells the lexer to skip any whitespace it encounters by
matching it and then calling itself recursively to continue.  (Note that
\textcd{lex} is an example of an identifier introduced implicitly
when the snippet is copied.)  ML-ULex\footnote{The lexer generator
(compatible with ML-Lex) that Standard ML of New Jersey uses.}~\cite{turon+:ml-lpt} converts the line into the
Standard ML code:

\codebegin
fun yyAction0 (strm, lastMatch : yymatch) =
   (yystrm := strm; ( lex () ))
\codeend

This output code already is not very easy to read.  However, the
problem is greatly exacerbated by the familiar phenomenon in typed
functional languages that type checkers are often bad at identifying
the true source of a type error.  Suppose we introduce an error into
the specification by omitting the argument to \textcd{lex}:

\codebegin
{whitespace}+ => ( lex );
\codeend

We now obtain\footnote{Using Standard ML of New Jersey v100.68.} several pages of error messages looking like:

\bigcodebegin
foo.lex.sml:1526.25-1526.70 Error: operator and
operand don't agree [tycon mismatch]
\z
 operator domain: yyInput.stream * action * yymatch
 operand:         yyInput.stream * 
                  (yyInput.stream * yymatch -> unit 
                  -> (?.svalue,int) ?.token)
                  * yymatch
 in expression:
   yyMATCH (strm,yyAction0,yyNO_MATCH)
\bigcodeend

\noindent
or like:

\bigcodebegin
foo.lex.sml:1686.20-1692.47 Error: types of if
branches do not agree [tycon mismatch]
\z
 then branch: (?.svalue,int) ?.token
 else branch: unit -> (?.svalue,int) ?.token
 in expression:
   if inp = #"\\n"
   then yyQ38 (strm',lastMatch)
   else 
     if inp < #"\\n"
     then if inp = #"\\t" then yyQ37 (<exp>,<exp>)
                         else yyQ36 (<exp>,<exp>)
     else yyQ37 (strm',lastMatch)
\bigcodeend

\noindent
and none of the errors is anywhere near the copied snippet containing
the error.

\medskip
The problem is related to the issue of variable hygiene in macro
expansion~\cite{kohlbecker+:hygienic-macros}.  In both cases, the
programmer writes code (a lexer/parser action, or macro argument)
divorced from its ultimate context and then---after processing---that
code is dropped verbatim into its ultimate context.  In the setting of
macros, this sets the scene for variable capture to occur, which is
nearly always erroneous.  In lexer generators, variable capture often is
actually desired (consider the \textcd{lex} identifier), but, as
observed above, it is nevertheless difficult to reason about and
to debug.

Accordingly, we are interested in source-code generation in which all
code is type-checked in the same context in which it is written.  We
call this {\em hygienic source-code generation\/} by analogy to
hygienic macro expansion, which ensures the same thing for macros.

An obvious way to accomplish hygienic source-code generation is to have the tool
type-check every snippet before it assembles them into output code.
But, this approach is unattractive in practice, because it necessitates including
all the apparatus of parsing, elaboration, and type-checking as part
of a tool that does not otherwise need all that apparatus.

We propose a simpler and cleaner alternative: Rather than type-check
disembodied code in context, we dispense with disembodied code
altogether.  To accomplish this, the tool---rather than assembling
snippets of source code into a program---generates a functor that
abstracts over the code that used to reside in snippets.  The
programmer then applies the functor in order to instantiate the lexer
or parser with specific action implementations.

A third alternative, arguably more principled than ours, is to
implement the lexer/parser generator in a type-safe metaprogramming
language such as MetaML~\cite{taha+:metaml-multi-stage-explicit-annotations} or its cousins.  With such an
approach, as in ours, the action implementations would be type-checked
in context, without any need to duplicate compiler apparatus.
Furthermore, it
would remove the need to write the lexer/parser specification and
action implementations in two separate places, as our proposal requires.
On the other hand, this alternative requires one to use a special
programming language.  We want an approach compatible with
pre-existing, conventional functional programming languages,
specifically ML and Haskell.

Finally, in some problem domains one may consider avoiding generated
source code entirely.  For example, in parsing, some programmers find parser
combinators~\cite{frost+:nl-interp-lazy-functional,hutton:combinator-parsing}
to be a suitable or even preferable alternative to Yacc-like tools.
Nevertheless, many programmers prefer traditional LR parser generators
for various reasons including error reporting and recovery, and
ambiguity diagnostics.  In this work we take it as given that source-code
generation is preferred, for whatever reason.

\medskip
At first blush, our proposal might seem to replace one sort of disembodied
code with another.  This is true in a sense, but there is a key
difference.  The code in which the functor is applied is {\em ordinary
code,} submitted to an ordinary compiler.  That compiler then
type checks the action code (that formerly resided in snippets) in the
context in which it now appears, which is the functor's argument.

As a practical matter, each action becomes a distinct field of the
functor argument, and consequently each action is type-checked
independently, as desired.  The type of the functor is already known,
so an error in one action will not be misinterpreted as an error
in all the other actions.

\medskip

Employing this design, we have implemented a lexer generator, called
CM-Lex, and a parser generator, called CM-Yacc.  Each tool supports
Standard ML, OCaml, and Haskell.\footnote{The tool is implemented in
Standard ML.}  Both tools are
available on-line at:

\begin{center}
{\tt www.cs.cmu.edu/\tilde crary/cmtool/}
\end{center}

In the remainder of the paper we describe how the tools work, taking
the lexer generator as our primary example.

\section{Lexing Functor Generation}
\label{sec:lexer}

The following is a very simple CM-Lex specification:

\codebegin
sml
name LexerFun
alphabet 128
\z
function f : t =
   (seq 'a 'a) => aa
   (seq 'a (* 'b) 'c) => abc
\codeend

The specification's first line indicates that CM-Lex should generate
Standard ML code.
The next two lines indicate that CM-Lex should
produce a functor named \textcd{LexerFun}, and that it should
generate a 7-bit parser (any symbols outside the range $0 \ldots
127$ will be rejected automatically).

The remainder gives the specification of a lexing function named
\textcd{f}.  The function will return a value of type \textcd{t}, and
it is defined by two regular expressions.  Regular expressions are
given as S-expressions using the Scheme Shell's SRE
notation\footnote{Although SREs are less compact than some other
notations, we find their syntax is much easier to remember.}
\cite{shivers:sre}.

Thus, the first arm activates an action named \textcd{aa} when the
regular expression $aa$ is recognized.  The second activates an
action named \textcd{abc} when the regular expression $ab^*c$ is
recognized.

Observe that the specification contains no disembodied code.  The
actions are simply given names, which are instantiated when the
resulting functor is applied.

\medskip

From this specification, CM-Lex generates the following Standard ML
code:\footnote{We simplify here and in the following examples for the
sake of exposition.}

\newcommand{\otherstuff}{{\it \dots other stuff \dots}}
\newcommand{\implementation}{{\it \dots implementation \dots}}
\codebegin
functor LexerFun
  (structure Arg :
     sig
       type t
\z
       type info = { match : char list,
                     follow : char stream }
\z
       val aa : info -> t
       val abc : info -> t
     end)
  :>
  sig
    val f : char stream -> Arg.t
  end
= \implementation
\codeend

When the programmer calls the functor, he provides the type \textcd{t}
and the actions \textcd{aa} and \textcd{abc}, both of which produce a
\textcd{t} from a record of matching information.  The functor then
returns a lexing function \textcd{f}, which produces a \textcd{t} from
a stream of characters.

Although the programmer-supplied actions can have side effects, the
lexer itself is purely functional.  The input is processed using lazy
streams (the signature for which appears in Figure~\ref{fig:streams}).
Each action is given the portion of the stream that follows the
matched string as part of the matching information.

\begin{figure}
\fighead
\codebegin
signature STREAM =
  sig
    type 'a stream
    datatype 'a front =
       Nil
     | Cons of 'a * 'a stream
\z
    val front : 'a stream -> 'a front
    val lazy : (unit -> 'a front) -> 'a stream
  end
\codeend
\caption{Lazy Streams}
\label{fig:streams}
\figfoot
\end{figure}

\medskip

As an illustration of how the functor might be applied, the following
program processes an input stream, printing a message each time it
recognizes a string:

\codebegin
structure Lexer =
  LexerFun
  (structure Arg =
     struct
       type t = char stream
\z          
       type info = { match : char list,
                     follow : char stream }
\z
       fun aa ({follow, ...}:info) =
          ( print "matched aa\\n"; follow )
\z
       fun abc ({follow, ...}:info) =
          ( print "matched ab*c\\n"; follow )
     end)
\z
fun loop (strm : char stream) =
  (case front strm of
     Nil => ()
   | Cons _ => loop (Lexer.f strm))
\codeend

The function \textcd{Lexer.f} matches its argument against the two
regular expressions and calls the indicated action, each of which
prints a message and returns the remainder of the stream.

Observe that the implementations of the actions (the fields
\textcd{aa} and \textcd{abc} of the argument structure) are ordinary
ML code.
As one consequence, the action code faces the standard type checker.
Moreover, each action's required type is unambiguously given by
\textcd{LexerFun}'s signature and the type argument \textcd{t}, so
error identification is much more accurate.

For example, suppose we replace the \textcd{aa} action with an
erroneous implementation that fails to return the remainder of the
stream:

\codebegin
fun aa ({follow, ...}:info) =
  ( print "matched aa\\n" )
\codeend

The type checker is able to identify the source of the error precisely,
finding that \textcd{aa} has the type \textcd{unit} instead of~\textcd{t}:

\codebegin
example.sml:8.4-29.12 Error: value type in 
structure doesn't match signature spec
    name: aa
  spec:   ?.Arg.info -> ?.Arg.t
  actual: ?.Arg.info -> unit
\codeend

\subsection{An expanded specification}
\label{sec:expanded}

We may add a second function to the lexer by simply adding another
function specification:

\codebegin
function g : u =
  (or (seq 'b 'c) (seq 'b 'd)) => bcbd
  epsilon => error
\codeend

In the parlance of existing lexer generators, multiple
functions are typically referred to as multiple {\em start
conditions\/} or {\em start states,} but we find it easier to think
about them as distinct functions that might or might not share some
actions.  In this case, the function \textcd{g} is specified to return
a value of type \textcd{u}.  Since \textcd{u} might not be the same
type as \textcd{t}, \textcd{g} cannot share any actions with
\textcd{f}.

The first arm activates an action named \textcd{bcbd} when the regular
expression $bc + bd$ is recognized.  The second arm activates an
action named \textcd{error} when the empty string is recognized.  Like
other lexer generators, CM-Lex prefers the longest possible
match, so an epsilon arm will only be used when the input string fails
to match any other arm.  Thus, the latter arm serves as an error
handler.\footnote{In contrast, the specification for \textcd{f} was
inexhaustive, so CM-Lex added a default error handler that raises an
exception.}

From the expanded specification, CM-Lex generates the functor:

\codebegin
functor LexerFun
  (structure Arg :
     sig
       type t
       type u
\z
       type info = { match : char list,
                     follow : char stream }
\z
       val aa : info -> t
       val abc : info -> t
       val bcbd : info -> u
       val error : info -> u
     end)
  :>
  sig
    val f : char stream -> Arg.t
    val g : char stream -> Arg.u
  end
= \implementation
\codeend

\section{Recursion in actions}

One important functionality for a lexer generator is the ability for
actions to invoke the lexer recursively.  For example, it is common
for a lexer, upon encountering whitespace, to skip the whitespace and
call itself recursively (as in the example in
Section~\ref{sec:intro}).\footnote{One way to accomplish this would be
to structure the lexer with a driver loop (such as the function
\textcd{loop} in Section~\ref{sec:lexer}), and for the action to
signal the driver loop to discard the action's result and recurse.
However, the earlier example notwithstanding, this is usually not the
preferred way to structure a lexer.}

This can be problematic because it requires recursion between the
lexer functor's argument and its result.

For example, consider a lexer that turns a stream of characters into a
stream of words.  The CM-Lex specification is:

\codebegin
sml
name WordsFun
alphabet 128
\z
set whitechar =
    (or 32 9 10)  /* space, tab, lf */
set letter = (range 'a 'z)
\z
function f : t =
   (+ whitechar) => whitespace
   (+ letter) => word
\codeend

\noindent
CM-Lex generates the functor:

\codebegin
functor WordsFun
  (structure Arg :
     sig
       type t
\z
       type info = { match : char list,
                     follow : char stream }
\z
       val whitespace : info -> t
       val word : info -> t
     end)
  :>
  sig
    val f : char stream -> Arg.t
  end
= \implementation
\codeend

A natural way\footnote{This simple implementation does not result in
the best behavior from the lazy streams, because forcing the output
stream causes the lexer to examine more of the input stream than
is necessary to determine the output stream's first element.  We
illustrate a better way to manage laziness in
Appendix~\ref{apx:full-example}.  In any case, laziness is orthogonal
to the issue being discussed here.} to implement the desired
lexer would be with a recursive module definition:

\codebegin
structure rec Arg =
  struct
    type t = string stream
\z
    type info = { match : char list,
                  follow : char stream }
\z
    fun whitespace ({follow, ...}:info) =
      Words.f follow
\z
    fun word ({match, follow, ...}:info) =
      lazy
      (fn () => Cons (String.implode match,
                      Words.f follow))
  end)
\z
and Words = WordsFun (structure Arg = Arg)
\codeend

Unfortunately, recursive modules bring about a variety of thorny
technical
issues~\cite{crary+:recmod,russo:types-for-modules,dreyer:phd}.
Although some dialects of ML support recursive modules, Standard ML
does not.

As a workaround, CM-Lex provides recursive access to the lexer via a
self field passed to each action.  The \textcd{info} type is
extended with a field \textcd{self : self}, where the type
\textcd{self} is a record containing all of the lexing functions being
defined.  In this case:

\codebegin
type self = { f : char stream -> t }
\codeend

Using the \textcd{self}-augmented functor, we can implement the lexer as follows:

\codebegin
structure Words =
  WordsFun
  (structure Arg =
     struct
       type t = string stream
\z
       type self = { f : char stream -> t }
       type info = { match : char list,
                     follow : char stream,
                     self : self }
\z
       fun whitespace 
           ({match, follow, self, ...}:info) =
         #f self follow
\z
       fun word 
           ({match, follow, self, ...}:info) =
         lazy 
         (fn () => Cons (String.implode match,
                         #f self follow))
     end)
\codeend

\section{Parsing Functor Generation}

The parser generator, CM-Yacc, works in a similar fashion to CM-Lex.
A CM-Yacc specification for a simple arithmetic parser is:

\codebegin
sml
name ArithParseFun
\z
terminal NUMBER of t
terminal PLUS
terminal TIMES
terminal LPAREN
terminal RPAREN
\z
nonterminal Term : t =
  1:NUMBER => number_term
  1:Term PLUS 2:Term => plus_term
  1:Term TIMES 2:Term => times_term
  LPAREN 1:Term RPAREN => paren_term
\z
start Term
\codeend

The specification says that the functor should be named
\textcd{ArithParseFun}, and it declares five terminals, one of which
carries a value of type \textcd{t}.

The specification then declares one nonterminal called \textcd{Term},
indicates that a term carries a value of type \textcd{t}, and gives
four productions that produce terms.\footnote{This grammar is
ambiguous, resulting in shift-reduce conflicts.  The ambiguity can be
resolved in either of the standard manners: by specifying operator
precedences, or by refactoring the grammar.}  Numbers are attached to
the symbols on the left-hand-side of a production that carry values
that should be passed to the production's action.  The number itself
indicates the order in which values should be passed.  Thus
\textcd{plus\_term} is passed a pair containing the first and third
symbols' values.

The final line specifies that the start symbol is \textcd{Term}.

\smallskip
From this specification, CM-Yacc generates the following Standard ML
code:

\codebegin
functor ArithParseFun
  (structure Arg :
     sig
       type t
\z
       val number_term : t -> t
       val plus_term : t * t -> t
       val times_term : t * t -> t
       val paren_term : t -> t
\z
       datatype terminal =
          NUMBER of t
        | PLUS
        | TIMES
        | LPAREN
        | RPAREN
\z
       val error : terminal stream -> exn
     end)
  :>
  sig
    val parse : Arg.terminal stream -> Arg.t
  end
= \implementation
\codeend

As before, the programmer supplies the type \textcd{t} and the
actions.  (The actions need not be passed a self argument, because
parser actions do not commonly need to reinvoke the parser.)  He also
supplies the \textcd{terminal} datatype and an error action, the
latter of which takes the terminal stream at which a syntax error is
detected and returns an exception for the parser to raise.  For
example:

\codebegin
datatype terminal =
   NUMBER of t
 | PLUS
 | TIMES
 | LPAREN
 | RPAREN
\z
structure Parser =
  ArithParseFun
  (structure Arg =
     struct
       type t = int
\z
       fun number_term x = x
       fun plus_term (x, y) = x + y
       fun times_term (x, y) = x * y
       fun paren_term x = x
\z
       datatype terminal = datatype terminal
\z
       fun error _ = Fail "syntax error"
     end)
\codeend

\noindent
Then our parser is $\mathcd{Parser.parse} : \mathcd{terminal -> int}$.

Note that we use datatype copying (a little-known feature of Standard
ML) to copy the \textcd{terminal} datatype into the \textcd{Arg}
structure.  If the datatype were defined within the \textcd{Arg}
structure, there would be no way to use it outside.  OCaml does not
support datatype copying, but one can get the same effect by including
a module that contains the datatype.

\section{Functors in Haskell}

In broad strokes the Haskell versions of CM-Lex and CM-Yacc are
similar to the ML versions.  In one regard, they are simpler: In
Haskell all definitions are mutually recursive, so no special
functionality is required to allow lexer actions to reinvoke the
lexer.

However, Haskell does not support functors, the central mechanism we
exploit here.  Instead, we built an ersatz functor from polymorphic
functions.

Recall the CM-Lex specification from Section~\ref{sec:expanded},
reprised in Figure~\ref{fig:reprise}.  From that specification, CM-Lex
generates a module (in the Haskell sense) named \textcd{LexerFun} with
the following exports:

\begin{figure}
\fighead
\codebegin
sml
name LexerFun
alphabet 128
\z
function f : t =
   (seq 'a 'a) => aa
   (seq 'a (* 'b) 'c) => abc
\z
function g : u =
  (or (seq 'b 'c) (seq 'b 'd)) => bcbd
  epsilon => error
\codeend

\centerline{\dots became \dots}

\codebegin
functor LexerFun
  (structure Arg :
     sig
       type t
       type u
\z
       type info = { match : char list,
                     follow : char stream }
\z
       val aa : info -> t
       val abc : info -> t
       val bcbd : info -> u
       val error : info -> u
     end)
  :>
  sig
    val f : char stream -> Arg.t
    val g : char stream -> Arg.u
  end
= \implementation
\codeend
\caption{The example from Section~\ref{sec:expanded}}
\label{fig:reprise}
\figfoot
\end{figure}

\codebegin
data LexInfo =
   LexInfo
   { match :: [Char],
     follow :: [Char] }
\z
data Arg t u =
  Arg { t :: Proxy t,
        u :: Proxy u,
\z
        aa :: LexInfo -> t,
        abc :: LexInfo -> t,
        bcbd :: LexInfo -> u,
        error :: LexInfo -> u }
\z
f :: Arg t u -> [Char] -> t
g :: Arg t u -> [Char] -> u
\codeend

Compare this with the ML version, also reprised in
Figure~\ref{fig:reprise}.  The type \textcd{Arg} represents the
argument to the functor.  It contains implementations for the four
actions \textcd{aa}, \textcd{abc}, \textcd{bcbc}, and \textcd{error}.

It also contains implementations for the two types \textcd{t} and
\textcd{u}.  Haskell does not support type fields like an ML
structure, but we can get a similar effect by including proxy fields
with the types \textcd{Proxy t} and \textcd{Proxy u}.  The programmer then fills
them in with the term \textcd{Proxy :: Proxy T} for
some~\textcd{T}.\footnote{Alternatively, one could give the proxy
fields the bare types \textcd{t} and \textcd{u} and fill them in with
\textcd{undefined :: T}, but that approach would be more awkward
in the monadic
case in which we also need to specify a monad.  A monad has kind
\cd{* -> *} and therefore does not have elements.}

{\tt Proxy}~\cite{haskell-proxy} is a type constructor in the Haskell standard library
that is designed for this sort of use.  For any type constructor {\tt
C}, the type \textcd{Proxy C} has a single data constructor {\tt Proxy}.
The {\tt Proxy} type constructor is poly-kinded, so {\tt C} need not
have kind \textcd{*}.

An alternative would be to leave out the type fields altogether and
allow type inference to fill them automatically.  We believe it would
be a misstep to do so.  The type implementations are critical
documentation that should be given explicitly in the program.
Moreover, leaving out the type implementations would bring back the
possibility that the type checker would misattribute the source of a
type error.

The functor's output is factored into two separate polymorphic
functions that each take the functor argument as an argument.  Since
the type arguments \textcd{t} and \textcd{u} are propagated to the
result types of the lexing functions, they must also appear as
explicit parameters of the type \textcd{Arg}.

\medskip

The Haskell version of CM-Yacc builds an ersatz functor in a similar
fashion.  However, while the ML version specified the terminal type
as an input to the parser functor, there is no way to specify a
datatype as an input to an ersatz functor.  Instead, the parsing
module defines the terminal datatype and passes it out.

In the example above, CM-Lex was used in purely functional mode.
Consequently, the input stream was simply a character list, since
Haskell lists are lazy already.  Alternatively, CM-Lex and CM-Yacc can
be directed to generate monadic code, which allows the lexer or parser
to deal with side effects, either in the generation of the input stream
({\em e.g.,} input read from a file) or in the actions.
Doing so incurs some complications --- it is important that the input
stream be memoizing and not every monad is capable of supporting the
proper sort of memoization\footnote{Monads such as \textcd{IO} and
\textcd{ST} that support references also support memoization, and
\textcd{Identity} supports it trivially (since no memoization need
ever be done), but most others do not.} --- but these complications
are orthogonal to the functor mechanism discussed here and are beyond
the scope of this paper.

\section{Conclusion}

We argue that functor generation is a cleaner mechanism for
source-code-generating tools than assembling snippets of disembodied
code.  The resulting functor makes no demands on the surrounding code
(other than a few standard libraries), and so it is guaranteed to type
check.\footnote{More precisely, it is guaranteed to type check in an
initial context containing standard libraries and other module
definitions.  Unfortunately, Standard ML does not quite enjoy the
weakening property, so the resulting functor is not guaranteed to type
check in {\em any\/} context.  Pollution of the namespace with
datatype constructors and/or infix declarations for identifiers that
are used within the generated functor will prevent it from parsing
correctly.  This is one reason why it is considered good practice in
SML for all code to reside within modules.} The programmer
never need look at the generated code.

In contrast, with a snippet-assembling tool, an error in any snippet
will --- even in the best case --- require the programmer to look at
generated code containing the snippet.  More commonly, the programmer
will need to look at lots of generated code having nothing to do with
the erroneous snippet.

We have demonstrated the technique for lexer and parser generation,
but there do not seem to be any limitations that would preclude its
use for any other application of source-code generation.

\appendix
\section{A Full Example}
\label{apx:full-example}

As a more realistic example, we implement a calculator that processes
an input stream and returns its value.  For simplicity, the calculator
stops at the first illegal character (which might be the end of the
stream).  The lexer specification is:

\codebegin
sml
name CalcLexFun
alphabet 128
\z
set digit = (range '0 '9)
set whitechar =
  (or 32 9 10)  /* space, tab, lf */
\z
function lex : t =
  (+ digit) => number
  '+ => plus
  '* => times
  '( => lparen
  ') => rparen
  (+ whitechar) => whitespace
\z
  /* Stop at the first illegal character */
  epsilon => eof
\codeend

\noindent
which generates:

\codebegin
functor CalcLexFun
  (structure Arg :
     sig
       type t
\z
       type self = { lex : char stream -> t }
       type info = { match : char list,
                     follow : char stream,
                     self : self }
\z
       val number : info -> t
       val plus : info -> t
       val times : info -> t
       val lparen : info -> t
       val rparen : info -> t
       val whitespace : info -> t
       val eof : info -> t
     end)
  :>
  sig
    val lex : char stream -> Arg.t
  end
= \implementation
\codeend

\noindent
and the parser specification is:

\codebegin
sml
name CalcParseFun
\z
terminal NUMBER of t
terminal PLUS
terminal TIMES
terminal LPAREN
terminal RPAREN
\z
nonterminal Atom : t =
  1:NUMBER => number_atom
  LPAREN 1:Term RPAREN => paren_atom
\z
nonterminal Factor : t =
  1:Atom => atom_factor
  1:Atom TIMES 2:Factor => times_factor
\z
nonterminal Term : t =
  1:Factor => factor_term
  1:Factor PLUS 2:Term => plus_term
\z
start Term
\codeend

\noindent
which generates:

\codebegin
functor CalcParseFun
  (structure Arg :
     sig
       type t
\z
       val number_atom : t -> t
       val paren_atom : t -> t
       val atom_factor : t -> t
       val times_factor : t * t -> t
       val factor_term : t -> t
       val plus_term : t * t -> t
\z
       datatype terminal =
          NUMBER of t
        | PLUS
        | TIMES
        | LPAREN
        | RPAREN
\z
       val error : terminal stream -> exn
     end)
  :>
  sig
    val parse : Arg.terminal stream -> Arg.t
  end
= \implementation
\codeend

\noindent
We then assemble the calculator as follows:

\bigcodebegin
structure Calculator
  :> sig
       val calc : char stream -> int
     end =
  struct
    datatype terminal =
       NUMBER of int
     | PLUS
     | TIMES
     | LPAREN
     | RPAREN
\z
    structure Lexer =
      CalcLexFun
      (structure Arg =
         struct
           type t = terminal front
\z
           type self = { lex : char stream -> t }
           type info = { match : char list,
                         follow : char stream,
                         self : self }
\z
           fun number
               ({ match, follow, self }:info) =
             Cons (NUMBER 
                     (Option.valOf 
                        (Int.fromString 
                           (String.implode match))),
                   lazy (fn () => #lex self follow))
\z
           fun simple terminal
               ({ follow, self, ... }:info) =
             Cons (terminal, 
                   lazy (fn () => #lex self follow))
\z
           val plus = simple PLUS
           val times = simple TIMES
           val lparen = simple LPAREN
           val rparen = simple RPAREN
\z
           fun whitespace 
               ({ follow, self, ... }:info) =
             #lex self follow
\z
           fun eof _ = Nil
         end)
\z
    structure Parser =
      CalcParseFun
      (structure Arg =
         struct
           type t = int
\z
           fun id x = x
\z
           val number_atom = id
           val paren_atom = id
           val atom_factor = id
           fun times_factor (x, y) = x * y
           val factor_term = id
           fun plus_term (x, y) = x + y
\z
           datatype terminal = datatype terminal
\z
           fun error _ = Fail "syntax error"
         end)
\z
    fun calc strm =
      Parser.parse 
      (lazy (fn () => Lexer.lex strm))
  end
\bigcodeend

\bibliographystyle{plain}
\bibliography{c:/crary/crary}

\end{document}